\documentclass[a4paper]{article}

\usepackage{INTERSPEECH2022}
\usepackage{amssymb}
\usepackage{makecell}
\usepackage{bbding}
\usepackage{bm} 
\usepackage{multirow}
\usepackage{array}
\usepackage{booktabs}
\usepackage{bbold}
\usepackage{arydshln}
\usepackage{url}
\title{SP-SEDT: Self-supervised Pre-training for Sound Event Detection Transformer}
\name{Zhirong Ye$^{1,2}$, Xiangdong Wang$^{1}$, Hong Liu$^1$, Yueliang Qian$^1$, Rui Tao$^3$, Long Yan$^3$, Kazushige Ouchi$^3$}
\address{$^1$Beijing Key Laboratory of Mobile Computing and Pervasive Device, \\Institute of Computing Technology, Chinese Academy of Sciences, Beijing, China\\
$^2$University of Chinese Academy of Sciences, Beijing, China\\
$^3$Toshiba China R\&D Center, Beijing, China}
\email{\{yezhirong19s, xdwang, hliu, ylqian\}@ict.ac.cn, \{taorui, yanlong\}@toshiba.com.cn, kazushige.ouchi@toshiba.co.jp}

\begin{document}

\maketitle
\begin{abstract}
  Recently, an event-based end-to-end model (SEDT) has been proposed for sound event detection (SED) and achieves competitive performance. However, compared with the frame-based model, it requires more training data with temporal annotations to improve the localization ability. Synthetic data is an alternative, but it suffers from a great domain gap with real recordings. Inspired by the great success of UP-DETR in object detection, we propose to self-supervisedly pre-train SEDT (SP-SEDT) by detecting random patches (only cropped along the time axis). Experiments on the DCASE2019 task4 dataset show the proposed SP-SEDT can outperform fine-tuned frame-based model. The ablation study is also conducted to investigate the impact of different loss functions and patch size.
\end{abstract}
\begin{figure*}[!t]
\centering
\includegraphics[width=7.1in]{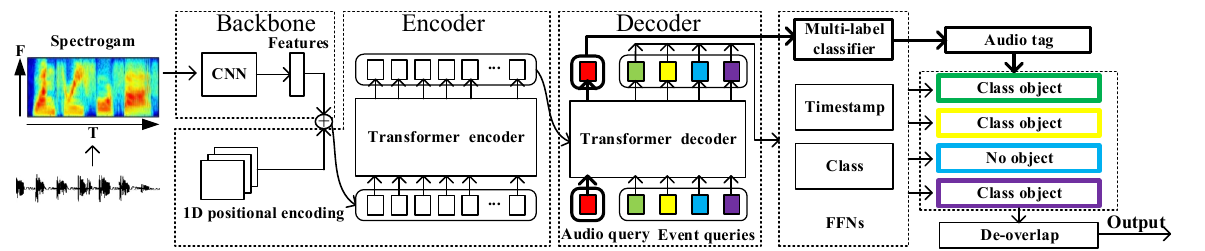}
\caption{Overview of Sound Event Detection Transformer.}
\label{sedt}
\end{figure*}

\section{Introduction}
\label{sec:intro}
Sound event detection (SED) aims to recognize sound events and identify their corresponding onset and offset in audio. SED plays an important role in many applications, such as sound-based surveillance \cite{crocco2016audio}, video indexing \cite{hershey2017cnn}, and so on.

Nowadays, deep learning models have become the mainstream approach for SED. Most of them formulate SED as a sequence-to-sequence classification problem, which predicts the event categories for each frame and then aggregates frame-level predictions  via post-processing to determine event boundaries, thus obtaining final event-level results \cite{zhang2015robust,turpault2019sound,miyazaki2020weakly}. However, post-processing including binarization by thresholding and median filtering with fixed or adaptive window sizes \cite{miyazaki2020weakly,DCASE2019ICT} brings many class-dependent hype-parameters to tune, and the tuning of such parameters may be dependent on a development dataset and inclined to overfitting in some cases. Moreover, under the conventional scheme, the model is optimized to achieve the best frame-level predictions instead of the best event-level results. The inconsistency between model training and evaluation may cause the model to focus more on local (frame-level) information and pay less attention to global and event-level characteristics, thus casting a restriction on the SED performance. To overcome the shortcomings of frame-based models, in our earlier work \cite{ye2021sound}, the object-based prediction mechanism of Detection Transformer (DETR) \cite{carion2020end} in the field of image object detection has been borrowed to build an event-based model for SED with transformer (SEDT), which outputs event-level predictions (including categories and boundaries) directly and achieves competitive performance to conventional methods. The event-based model pursues the same goal with the detection task, namely the best event-level predictions, making the model better adapt to the SED task. Besides, the model is capable of learning various event characteristics such as duration, which can only be inserted into the frame-based model by the hyperparameters in the post-processing stage.

However, compared with the conventional two-stage detection systems, the event-based model is more data-hungry \cite{zhigang2020updetr}. Unfortunately, the shortage of training data is more severe for SED than for other tasks such as image object detection, speech recognition or acoustic scene classification. Actually, due to the high cost of annotated audio clips with temporal boundaries of events, weakly supervised learning \cite{miyazaki2020weakly,lin2020guided, 2018Adaptive} has been applied to train SED models with weakly labeled data only indicating the event categories occurring in audio clips. Synthetic data \cite{salamon2017scaper} has also been utilized, which is often cooperated with domain adaption techniques \cite{wei2020A-CRNN, huang2020learning} to narrow the feature gap between real recordings and synthetic data.

Self-supervised learning has emerged as a paradigm to tackle with low-resource model training and has achieved remarkable progress in many fields, such as natural language processing \cite{DevlinCLT19bert}, computer vision \cite{HeZRS16resnet,CaronMMGBJ20swav} and automatic speech recognition \cite{BaevskiZMA20wav2vec}. Self-supervised pre-training has also been introduced into the audio community. Deshmukh et al. \cite{soham2021weakly} propose to jointly train the feature encoder with the SED task and an auxiliary self-supervised spectrogram reconstruction task. In \cite{lee2019labelefficient}, multiple self-supervised tasks are implemented to train a feature extractor for audio classification tasks. Almost all existing pre-training methods in the audio fields aim to learn better signal representation, which is conducive to classification or recognition tasks. However, the object-based model has two subtasks: classification and localization, where the latter is not a classification task as the model outputs temporal boundaries. Therefore, existing pre-training methods cannot be applied to object-based detection models directly. To improve the localization ability by pre-training, Dai et al. \cite{zhigang2020updetr} propose UP-DETR, which pre-trains DETR by detecting randomly cropped patches, and achieves great improvement on image object detection. But such spatial-aware pre-training has not been explored in SED.

In this paper, inspired by UP-DETR, we propose to our knowledge the first self-supervised learning approach to pre-train the event-based model SEDT to equip it with better localization ability, which is named as Self-supervised Pre-training SEDT (SP-SEDT). Specifically, we crop the spectrogram along the time axis to obtain random-sized patches at random locations, and then pre-train the model to predict corresponding temporal boundaries of the patches. To preserve the category information in SP-SEDT, classification loss and patch feature reconstruction loss are also added to optimize SP-SEDT. Experimental results on the DCASE 2019 Task4 dataset prove the effectiveness of SP-SEDT. And the ablation study shows that the category-information-related losses help to preserve the classification ability during localization pre-training. Besides, the random patch size can equip SP-SEDT with better generalization on sound events with varied duration. The code is available at \url{https://github.com/Anaesthesiaye/sound_event_detection_transformer.git}. 

\section{Sound Event Detection Transformer}
In this section, we describe the main idea of Sound Event Detection Transformer (SEDT) briefly, based on which our self-supervised pre-training method is proposed. More details of SEDT can be found in \cite{ye2021sound}.
\label{sec:SEDT}
\subsection{Model architecture}
An overview of SEDT is illustrated in Figure \ref{sedt}. The model contains three main components, including a backbone, an encoder-decoder transformer, and prediction feed-forward networks (FFNs). SEDT represents each sound event as a vector $y_i=(c_i,b_i)$, where $c_i$ is the class label and $b_i=(m_i,l_i)$ denotes the temporal boundary containing normalized event center $m_i$ and duration $l_i$. Such event representation allows SEDT to output event-level results directly.

The backbone and encoder are adopted for feature extraction. Given the spectrogram of an audio clip, the backbone is utilized to extract its feature map $z_0$. Then the sum of a positional encoding $P$ and the feature map $z_0$ is flattened on the time and frequency axis, resulting in a $d\times TF$ feature map $z$, which is fed into the transformer encoder \cite{vaswani2017attention} for further feature extraction. The cascade of CNN and transformer encoder ensures that the model can consider both local and global features.

The decoder is used to generate event and clip representations from the encoded feature. SEDT models SED as a set prediction problem and assumes that each event is independent. Therefore, the standard auto-regressive decoding mechanism adopted in machine translation task is discarded. Instead, the decoder takes $N$ learned embeddings as input, named as \emph{event queries}, and output $N$ event representations in parallel, where $N$ is a hyper-parameter and larger than the typical number of events in an audio clip. An extra \emph{audio query} is inserted along with the event queries and is fed into the decoder together. The audio query aggregates clip-level representation for audio tagging task and ensures that enough category information is utilized during the decoding of event representation.  

Prediction FFNs are adopted to transform the event and clip representation from the decoder into event detection results $\hat{y}_i=(\hat{c}_i,\hat{b}_i)$ and audio tags $y_\text{TAG}$. 


\subsection{Loss function}
\textbf{Event-level loss:}
To compute the event-level loss, the matching between target and predicted events is needed, which can be obtained by the Hungarian algorithm \cite{ye2021sound}. The unmatched predictions will be labeled as ``empty''. The loss function includes \emph{location loss} and \emph{classification loss}:
\begin{equation}
\mathcal{L_{\text{event}}}=\mathcal{L}_{\text{loc}}+\mathcal{L}_{\text{c}} \label{eq4}
\end{equation}
The \emph{location loss} is calculated for predictions that are not matched to ``empty" events ($c_{i} \neq \emptyset$). It is a linear combination of the L1 norm and IOU loss between the target and predicted location vector:
\begin{align}
    \mathcal{L}_{\text{loc}}=\sum_{i}^{N}\mathbb{1}_{\left\{c_{i} \neq \emptyset \right\}} \left(\lambda_{\text{IOU}}\mathcal{L}_{\text{IOU}}(b_i,\hat{b}_{\hat{\sigma}(i)})+\lambda_{\text{L1}}\parallel b_i-\hat{b}_{\hat{\sigma}(i)} \parallel_1 \right) 
\end{align}
where $\lambda_{\text{IOU}}, \lambda_{\text{L1}}\in \mathbb{R}$ are hyperparameters, $\hat{\sigma}$ is the assignment given by the matching process, and $N$ is the number of predictions. The \emph{classification loss} is the cross-entropy between the labels and the predictions:
\begin{equation}
\mathcal{L}_{\text{c}}=\frac{1}{N}\sum_{i=1}^N -\operatorname{log}\hat{p}_{\hat{\sigma}(i)}(c_i) \label{eq6}
\end{equation}
\textbf{Clip-level loss}: The audio tagging loss is calculated as the binary cross-entropy between the clip-level class label $l_\text{TAG}$ and predicted audio tagging $y_\text{TAG}$:
\begin{equation}
\mathcal{L}_{\operatorname{at}}=\operatorname{BCE}(\boldsymbol{l}_{\text{TAG}},\boldsymbol{y}_{\text{TAG}})
\end{equation}
SEDT is optimized by minimizing both event-level and clip-level losses, while for weakly labeled data, only clip-level loss is calculated:
\begin{align}
     \mathcal{L}_{\text{strong}} &=\mathcal{L_{\text{event}}}+\lambda_{at}\mathcal{L}_{\text{at}}\\ \nonumber
    \mathcal{L}_{\text{weak}} & =\lambda_{at}\mathcal{L}_{\text{at}}   
\end{align}
where $\lambda_{at}$ is a hyperparameter.
\section{Self-supervised Pre-training SEDT}
In this section, we describe the training of SP-SEDT in detail, including pre-training and fine-tuning.
\begin{figure}[!t]
\centering
\includegraphics[width=3.3in]{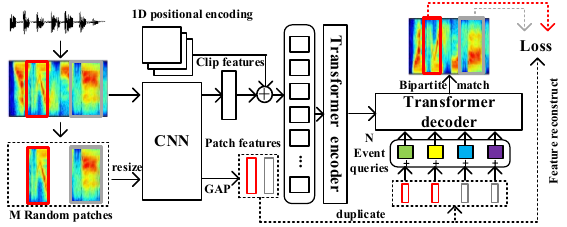}
\caption{Overview of Self-supervised Pre-training SEDT, where $N=4, M=2$.}
\label{SP-SEDT}
\end{figure}
\subsection{Random patch detection}
Without annotated event boundaries, SP-SEDT is pre-trained to detect randomly cropped patches (only cropped along the time axis with both random size and position). As depicted in Figure \ref{SP-SEDT}, SP-SEDT shares almost the same model architecture as SEDT except for the audio query branch. The audio query branch of SEDT is utilized for audio tagging, which requires weakly-labeled data to train and cannot work on unlabeled data. 

The pre-training procedure is also shown in Figure \ref{SP-SEDT}. Considering that there is usually more than one event occurring in an audio clip, during pre-training, $M$ patches are randomly cropped from the Mel-spectrogram. Each patch is fed into the backbone and a global average pooling (GAP) module to obtain patch feature $p$. In SEDT, the learned event queries have different detection patterns focusing on different areas and sizes related to different sound events \cite{carion2020end}. However, during pre-training in SP-SEDT, the event queries would fail to learn those latent properties of patches, which are cropped randomly. To preserve the specialization of queries, the information of patches needs to be injected into the queries to guide the query slot to detect the corresponding patch. Specifically, for an SP-SEDT model with $M$ randomly cropped patches and $N$ event queries, each patch feature $p$ will be duplicated and added to $\frac{N}{M}$ queries to obtain the final input of decoder. It is noteworthy that although the patch feature is fed into the decoder, it doesn't contain the position information of the patch in the original spectrogram. The SP-SEDT has to locate the patch by itself.

For each patch, there are three subtasks for SP-SEDT to complete, including classification, localization and feature reconstruction. For classification, a candidate patch is classified as ``patch'' ($c_i=1$) or ``background'' ($c_i=0$). The localization subtask aims to predict temporal boundaries. As for feature reconstruction, it is introduced to balance classification and localization during pre-training. We hope the event representation $\hat{p}$ from the decoder can contain as much category information as the patch feature $p$ extracted by the backbone from the spectrogram. Hence, the feature reconstruction loss is formulated as the mean squared error between the L2 normalized $p$ and $\hat{p}_{\hat{\sigma}(i)}$, where $\hat{\sigma}(i)$ is obtained by bipartite match:
\begin{align}
\mathcal{L}_{\text{rec}}=\parallel \frac{p_i}{\parallel p_i\parallel_2^2}-\frac{\hat{p}_{\hat{\sigma}(i)}}{\parallel \hat{p}_{\hat{\sigma}(i)}\parallel_2^2} \parallel_2^2
\end{align}
The loss function of classification, localization and feature reconstruction are linearly combined to jointly optimize SP-SEDT, where localization and reconstruction loss will only be computed for predictions labeled as ``patch'' ($c_{i}=1$):
\begin{equation}
\mathcal{L}=\mathcal{L}_{\text{c}}+\mathbb{1}_{\left\{ c_{i}=1 \right\}}(\mathcal{L}_{\text{loc}} +\mathcal{L}_{\text{rec}})
\end{equation}

\subsection{Backbone pre-training}
The random patch detection pre-training optimizes the model towards better localization ability without considering the classification ability. To avoid the model locates better at the cost of reduced recognition performance, we attempt to preserve the category information extraction ability of the model by freezing the parameters of the backbone during pre-training. Before pre-training, we need to guarantee the backbone can extract audio features depicting the spectrogram well. Existing self-supervised methods designed for better audio representation can be adopted to pre-train the backbone. Since this paper only focuses on the localization pre-training, to simplify the training process, we tuned the backbone (ResNet-50 pre-trained on ImageNet) on the weakly labeled data via audio tagging task.

\subsection{Fine-tuning}
After pre-training, the parameters of backbone, transformer and the prediction FFNs for temporal boundary regression of the SP-SEDT are utilized to initialize the SEDT model, while other modules, including the audio query branch and the prediction FFNs for classification, are initialized randomly.  
\section{EXPERIMENTS}
\label{sec:pagestyle}

\subsection{Experimental setup}

The experiments are conducted on the DCASE2019 task4 dataset, which consists of 10 sound event categories. It can be divided into a training set and a validation set (1168 clips), the training set contains a weakly-labeled subset (1578 clips), a strongly-labeled synthetic subset (2045 clips) and an unlabeled subset (14412 clips). SP-SEDT are pre-trained on the unlabeled subset and the DCASE2018 task5 development dataset (72984 clips), and the backbone is tuned on the weakly-labeled subset before pre-training. Overall, in our experiments, we first train the backbone by the audio tagging task using the weakly labeled data, then pre-train the SP-SEDT on the unlabeled subset and the DCASE2018 task5 development set, finally fine-tune it with the weakly-labeled and synthetic subset, and report the metric on the validation set. The DCASE2019-2021 task4 dataset differs only in the synthetic subset.

As for the baselines, we use CRNN and the Transformer-based model \cite{miyazaki2020weakly} referred to as ``CTrans'' in this paper. The baselines are the vanilla models of the DCASE2020 1st system which replaces RNN or Transformer encoder with Conformer encoder. The baselines output frame-level results and are then binarized and smoothed by class-adaptive thresholds and median filters. The baselines are trained on the weakly-labeled and synthetic subset of the DCASE2019 task4 dataset, the same as the SP-SEDT fine-tuning data.

We calculate the event-based measure (200ms collars on both onsets and offsets) and segment-based measure with a segment length of 1s by the sed\_eval package \cite{mesaros2016metrics} to evaluate the detection performance of models. The clip-level macro F1 score is also reported to measure the audio tagging performance.

\subsection{Results}
\begin{table}[ht!]
    \centering
    \caption{Performance on DCASE2019 Task4 validation set, where Eb, Sb, and At denotes Event-based, Segment-based, and Audio tagging macro $F1$ respectively.}
   \begin{tabular}{cccc}
    \toprule
         Model & Eb$[\%]$ & Sb$[\%]$ & At$[\%]$\\
         \midrule
         CRNN\cite{miyazaki2020weakly}& 30.61 & 62.21 & - \\
         CTrans($E=3$)\cite{miyazaki2020weakly} & 34.27 & $\bm{65.07}$ & - \\
         CTrans($E=6$)\cite{miyazaki2020weakly} & 34.28 & 64.33 & - \\
         \midrule
         SEDT($E=3$) &31.39 & 58.12 & 65.09 \\
         SP-SEDT($E=3$) & 38.85  & 64.10 & $\bm{73.01}$ \\
         SEDT($E=6$) & 31.64 & 61.18 & 69.11  \\
         SP-SEDT($E=6$) & $\bm{39.03}$& $64.54$ & $72.29$ \\
         \bottomrule
    \end{tabular}
    \label{tab:dcase}
\end{table}
The results on the DCASE2019 task4 validation set are given in Table \ref{tab:dcase}. We experiment with two model configurations that all adopt 3 transformer decoder blocks but differ in the number of transformer encoder blocks $E$. SP-SEDT (E=6) achieves the best results in all metrics. Among them, the SED main evaluation metric, i.e. event-based $F1$ score, gets the highest improvement with an average of $7.4\%$, which demonstrates that the random patch detection pretext task can equip SEDT with better localization ability. Besides, the segment-based and audio tagging $F1$ scores also increase by an average of $4.7\%$ and $5.6\%$, respectively. Actually, without strongly labeled data, SEDT can only learn to detect temporal boundaries from the synthetic data, which has a large domain gap with real recordings. While the proposed pre-training is conducted on real recordings, the localization ability of SP-SEDT can be better generalized to detect events in real audios. 

The performance of SEDT does not exceed CTrans, the main reason is that CTrans searches for the optimal binarization threshold and median filtering window size for each event category on the validation set, but SP-SEDT can outperform all baseline models without using the information of the validation set.

From Table \ref{tab:dcase}, we can also observe that SP-SEDT can perform better with the increasing of encoder blocks, while the deeper encoder does not bring benefit to the baseline models. It suggests that the pre-training allows training a larger model on limited labeled data without the triggering of overfitting. 




\subsection{Ablation study}
\begin{table}[ht!]
    \centering
    \caption{Ablation study on loss functions and patch size, where Eb denotes Event-based macro $F1$.}
    \begin{tabular}{ccccccc}
    \toprule
         \makecell[c]{DCASE2018 \\task5 data} & $\mathcal{L}_{\text{loc}}$ & \makecell[c]{Random \\ patch size} &$\mathcal{L}_{\text{c}}$ & $\mathcal{L}_{\text{rec}}$ & Eb$[\%]$ \\
         \midrule
         \XSolidBrush&\checkmark & \checkmark & \checkmark &  \checkmark & 36.52\\
         \checkmark& \checkmark & \checkmark & \checkmark & \XSolidBrush & 30.49  \\
         \checkmark&\checkmark & \checkmark & \XSolidBrush &  \checkmark & 37.49 \\
         \checkmark&\checkmark & \XSolidBrush & \checkmark &  \checkmark & 35.42 \\
         \checkmark&\checkmark & \checkmark & \checkmark &  \checkmark & $\bm{38.85}$  \\
         \bottomrule
    \end{tabular}
    \label{tab:ablation}
\end{table}
To speed up the progress of the experiment, we conduct ablation studies using the SP-SEDT($E=3$). Table \ref{tab:ablation} shows the results of models trained with different combinations of loss functions. It can be seen that the reconstruction loss has a significant impact, and the classification loss also matters. They both aim to preserve the category information in SP-SEDT during pre-training, so that the localization ability can be improved without a reduction in classification performance. 

Since patches have to be resized to a unified dimension during training, to avoid resizing influencing the spectrogram information, we attempt to use a unified patch size inner one batch but vary it between batches and compare this training strategy with the resized random patches pre-training. The results are given in Table \ref{tab:ablation}. Surprisingly, SP-SEDT with resized random patches has better performance, which suggests time-warping on the spectrogram won't bring a negative impact. On the contrary, it can equip the model with better generalization ability.   
\subsection{Results for the DCASE Challenge Task}
\begin{table}[ht!]\footnotesize
    \centering
    \caption{System performance on the DCASE2019 Task4 validation set and public evaluation set, where ME denotes model ensembling, Eb denotes Event-based macro $F1$.}
    \begin{tabular}{ p{1cm}<{\centering} p{1.3cm}<{\centering} p{2cm}<{\centering}p{0.3cm}<{\centering}p{0.5cm}<{\centering}<{\centering}p{0.5cm} }
        \toprule
         \makecell{Rank} &\makecell{Model} & Semi-supervised & \multirow{2}{*}{ME} &  \multicolumn{2}{c}{Eb[\%]} \\
         (year) & (\#param) & method & & Val &  Eval$_p$ \\
         \midrule
          \makecell{1st \cite{Lin2019}\\(2019)}& \makecell{CNN\\(7M)} & Guided learning & \checkmark &  45.3 &  -  \\
          \cdashline{1-6}
          \makecell{1st \cite{Miyazaki2020}\\(2020)}&\makecell{Conformer\\(17M)} & Mean teacher &  \checkmark &  \textbf{50.6} & \textbf{55.7}  \\
          \cdashline{1-6}
          \makecell{3rd \cite{Ebbers2020}\\(2020)}&\makecell{CRNN\\(20M)} & Pseudo-labelling &  \checkmark & 48.3 & 50.9 \\
          \cdashline{1-6}
            \multirow{2}{*}{ours} & SP-SEDT & \multirow{2}{*}{Pseudo-labelling} &  \multirow{2}{*}{\XSolidBrush} & \multirow{2}{*}{49.3} & \multirow{2}{*}{52.3}   \\
            & (36M) & \\
          \bottomrule
    \end{tabular}
    \label{tab:systems}
\end{table}
Experiments above only use the weakly labeled and synthetic subsets to compare among different model structures. However, in the DCASE challenges, semi-supervised methods are adopted to train models with unlabeled data, and other techniques such as model ensembling and data augmentation are also utilized. To compare with the top systems of the DCASE task4, we construct a SP-SEDT-based system, which uses pseudo-labelling, a simple method, for semi-supervised learning and perform mixup \cite{mixup} and SpecAugment \cite{park19e_interspeech} for data augmentation. To be fair, the SP-SEDT is pretrained only on the DCASE2019 unlabeled data. Table \ref{tab:systems} gives the results on the validation and public evaluation set. As can be seen, our system can achieve competitive performance. Considering that model ensembling and semi-supervised method adaptive to SP-SEDT haven't been further explored, we think SP-SEDT is of great potential. 

Since SP-SEDT has to locate and recognize at the same time, it owns higher model complexity, while we argue this means it can be more adaptive to pre-training. In the future, we would pretrain SP-SEDT in AudioSet \cite{audioset}, a large-scale dataset of manually-annotated audio events, to full play the role of large models. Besides, our purpose is not entirely to surpass the SOTA system, but to demonstrate that SEDT, a new and promising model architecture, can also be improved through its unique pre-training mechanism to achieve better results.

\section{Conclusion}
In this paper, we proposed a self-supervised pre-training method for sound event detection transformer to improve its localization ability by unlabeled data. Experimental results show that the proposed method improves the performance of SEDT and outperforms the frame-based models on the DCASE2019 task4 dataset. The pre-training method also allows training a deeper model on limited labeled data to obtain further gains.


\bibliographystyle{IEEEtran}

\bibliography{mybib}


\end{document}